\documentclass{article}

\usepackage{amsfonts, mathrsfs, amssymb, amsmath, enumerate, bbm, bm, mathtools}
\usepackage{multirow, multicol}
\usepackage{algorithm, algpseudocode} 
\usepackage{xcolor}
\usepackage{subcaption}

\def\longtitle{Tree-Based Predictive Models for Noisy Input Data}
\def\shorttitle{\longtitle}
\newcommand\defeq{\mathrel{\overset{\makebox[0pt]{\mbox{\normalfont\tiny\sffamily def}}}{=}}}

\definecolor{my_blue}{RGB}{65, 105, 225}

\usepackage{arxiv}
\usepackage[utf8]{inputenc} 
\usepackage[T1]{fontenc}    
\usepackage{hyperref}       
\usepackage{url}            
\usepackage{booktabs}       
\usepackage{nicefrac}       
\usepackage{microtype}      
\usepackage{cleveref}       
\usepackage{graphicx}
\usepackage{natbib}
\usepackage{doi}

\title{\longtitle}
\date{\today}

\author{\AND \href{https://orcid.org/0000-0002-3570-6826}{\includegraphics[scale=0.06]{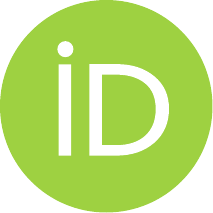}\hspace{1mm}Kevin McCoy} \\
    Department of Statistics\\
    Rice University\\
    Houston, TX 77005 \\
    \texttt{kmm12@rice.edu} \\
	\And
	\href{https://orcid.org/0000-0002-7827-1586}{\includegraphics[scale=0.06]{orcid.pdf}\hspace{1mm}Zachary Wooten} \\
    Department of Biostatistics\\
    St. Jude Children's Research Hospital\\
    Memphis, TN 38105\\
    \texttt{zachary.wooten@stjude.org} \\
	\And
	\href{https://orcid.org/0000-0003-3316-0468}{\includegraphics[scale=0.06]{orcid.pdf}\hspace{1mm}Christine B. Peterson\thanks{Author to whom correspondence should be addressed.}} \\
    Department of Statistics\\
    Rice University\\
    Houston, TX 77005 \\
    \texttt{cbp2@rice.edu} \\
}

\hypersetup{
pdftitle={\longtitle},
pdfsubject={stat.ML},
pdfauthor={Kevin McCoy, Zachary Wooten, Christine B. Peterson},
pdfkeywords={decision trees, Bayesian additive regression trees, measurement error, uncertainty quantification, latent variable model, Bayesian hierarchical modeling},
}

\begin{document}
\maketitle

\begin{abstract}\label{sec:abstract}
Measurement error is prevalent across all domains of scientific research where only imprecise observations, rather than the true underlying values, can be obtained. 
For example, estimates of human microbiome diversity are based on small samples from a much larger, generally unobserved system and reflect both sampling error and technical variation.
In high-noise settings like these, it becomes difficult to make accurate predictions and to summarize uncertainty. 
Methods have previously been proposed to accommodate measurement error in classic predictive models, such as linear regression.
However, relatively little work has been done to address measurement error in more complex and flexible models.  
Bayesian additive regression trees (BART), a Bayesian nonparametric model that sums the output of many decision trees, offers robust predictions with built-in uncertainty quantification. 
In this work, we propose measurement error BART (meBART), a novel extension to the BART model that directly incorporates measurement error in the independent variable(s). 
Through simulation studies, we show that in the presence of measurement error, our model enables more accurate parameter estimation, more robust uncertainty quantification, and superior predictive performance. We illustrate the utility of our proposed approach through two biomedical applications where the predictors of interest are subject to measurement error.
\end{abstract}

\keywords{decision trees \and Bayesian additive regression trees \and measurement error \and uncertainty quantification \and latent variable model \and Bayesian hierarchical modeling}

\section{Introduction}\label{sec:introduction}



Measurement error is widespread in many fields, including medicine, epidemiology, environmental science, economics, and survey sampling. Measurement error refers to the idea that a variable cannot be quantified with precision. Instead, only a rough surrogate for its value can be obtained. When a continuous predictor used as input to a statistical model is mismeasured, this phenomenon is referred to as errors-in-variables or errors-in-covariables 
\citep{gustafson2003measurement}. In many statistical analysis settings, particularly regression, we are concerned with estimating the relationship between a set of input variables and a response or outcome variable. 
Many statistical models allow for uncertainty in the outcome variable, but typically assume that the predictors are fixed values observed without noise. Measurement error models, in contrast, allow for measurement errors in the input variables of the prediction model. 

Measurement error may refer to random noise, sampling error, or variation associated with the measurement process itself \citep{yi_handbook_2021}. As an example, consider the task of estimating microbiome diversity based on metagenomic profiling of a patient stool sample.
Sampling error arises as the physical sample itself is a tiny snapshot of the much larger target population of bacteria inhabiting the gastrointestinal tract \citep{willis2019rarefaction}. 
Additional sources of measurement error include
technical noise due to differences in sample collection, handling, and sequencing \citep{vogtmann2017comparison}. Notably, even replicates obtained by splitting a single sample will not yield the same estimates of microbial abundance
\citep{ji2019quantifying}.
Finally, there may be naturally occurring variation from day to day due to changes in dietary intakes or the underlying dynamics of the system. 

In the current work, we demonstrate that ignoring measurement error in the input variables can have profound effects on inferential accuracy in predictive modeling. However, measurement error is typically disregarded and has not generally been considered in modern statistical learning approaches. In this work, we 
propose a novel approach for handling
measurement error in the predictors for Bayesian additive regression trees (BART) \citep{chipman_bart_2010}, a Bayesian nonparametric machine learning method with robust prediction performance on complex datasets \citep{hill_bayesian_2020}. 
Specifically, we represent the true unobserved feature values as latent variables in a Bayesian hierarchical model, and use these latent variables as inputs to the prediction model. 

Our paper is organized as follows.
In Section \ref{subsec:background}, we provide a brief background on measurement error models and  tree-based prediction methods. In Section \ref{sec:methods}, we introduce our proposed Bayesian hierarchical model, measurement error BART (meBART), including the prior formulation and posterior sampling approach. In Section \ref{sec:results}, we 
benchmark the performance of meBART against alternative approaches
in a variety of synthetic data settings. 
We apply meBART to two real-world data sets in Section \ref{sec:real_data}, where the predictors of interest, microbiome diversity and self-reported exercise activity, are subject to substantial measurement error. 
We interpret the results in Section \ref{sec:discussion} and make concluding remarks in Section \ref{sec:conclusion}.


\subsection{Background}\label{subsec:background}


\subsubsection{Measurement error models}

In a regression model, measurement errors can be present in the input variables, outcome variables, or both. 
In standard regression, a noise or error term is commonly included in the model to account for the uncertainty in the relationship between the input and output variables. For example, in the linear regression model $y=X\beta +\varepsilon$, the term $\varepsilon$ represents the error. This is slightly different than the concept of a mismeasured output variable $Y$, but in practice, the error term can capture both the model noise and measurement error in $Y$.
For this reason, measurement error literature exploring noise in the input variables is more common and will be the focus of this background.



The simplest example of a measurement error model is a univariate linear regression model that allows for noise in the quantification of the input variable \citep{yi_handbook_2021}. Consider a sequence of $n$ i.i.d.\ random observations $(x_i, y_i)$, where $i=1, \ldots, n$ indexes the observation. Here, $x_i$ and $y_i$ are scalars, and $\boldsymbol{x}$ and $\boldsymbol{y}$ denote the vectors of observations across samples. Instead of measuring $x_i$ directly, we measure a noisy surrogate $x_i^*$.  Assuming that the response variable $y_i$ depends on the true unobserved value $x_i$ and that the measurement noise $e_i$ is additive, we obtain the model:

\begin{equation}
\label{me_lr}
\begin{split}
    y_i = \beta_0& + \beta_x x_i + \varepsilon_i  \\
    x_i^* &= x_i + e_i.
\end{split}
\end{equation}

It is usually assumed that $\varepsilon_i$ is independent of  $x_i$ and has mean $0$ and variance $\sigma^2$. Similarly, it is commonly assumed that $e_i$ has mean $0$ and variance $\sigma_e^2$. Finally, $e_i$ is assumed to be independent of both $x_i$ and $\varepsilon_i$. This last assumption is referred to as nondifferential measurement error, and it follows that $x_i^*$ provides no extra information about $y_i$ if $x_i$ is known \citep{carroll_measurement_2006}. Modeling the noisy surrogate $x_i^*$ as a normal variable centered at the true value $x_i$ results in what is commonly known as the classical measurement error model. 

The impact of na\"ively using $x_i^*$ as $x_i$ in equation \eqref{me_lr} is twofold: (1) the estimator $\hat{\beta}_x^*$ is not a consistent estimator for the true slope $\beta_x$, and (2) the estimator $\hat{\beta}_x^*$ actually converges in probability to $\lambda\beta_x$, where $\lambda = \frac{\sigma^2_x}{\sigma^2_x+\sigma^2_e}$, called the reliability ratio \citep{buonaccorsi2010measurement}. Here, $\sigma^2_x$ is the variance of the random variable $X$. 
This attenuation when estimating $\mathbb{E}[Y|X^*]$ highlights why measurement error in the input variables can have a profound effect if ignored. 
In contrast, when measurement error is only present in the $Y$ variable, estimating $\mathbb{E}[Y^*|X]$ does not result in any bias, but instead only increases the uncertainty about the $(X, Y)$ relationship \citep{gustafson2003measurement}.

An alternative to the na\"ive approach in estimating $\mathbb{E}[Y|X]$ is regression calibration, which involves regressing $Y$ onto $\mathbb{E}[X|X^*]$. This method assumes that the true values of $\mu_x, \sigma_x^2$, and $\sigma_e^2$ are known and approximates $\mathbb{E}[X|X^*]$ by rescaling $\boldsymbol{x}^*$ into the scale of $\boldsymbol{x}$. However, this method is only an approximation, and if we suppose that $\varepsilon_i$, $e_i$, and $x_i$ are all normally distributed with unknown variance, the model would no longer be identifiable. Further assumptions can be made to rectify this problem. Two common solutions are to assume the ratio $\sigma_x^2/\sigma_e^2$ is known or to fix  $\lambda$, $\sigma^2$, or $\sigma^2_e$ to a known value \citep{yi_handbook_2021}. In this work, we take the second approach by assuming $\sigma^2_e$ to be known. In practice, this value can be estimated from an independent, `calibration' dataset. This assumption is fairly reasonable, as many sources of medical measurement error, for instance, come from devices with known variance. 
A rough estimate of $\sigma^2_e$ can also be obtained from repeated sampling within an individual within a short time window \citep{gustafson2003measurement}.


Measurement error models in the Bayesian framework treat the unobserved true latent values as parameters to be estimated. 
There are many advantages to taking a Bayesian approach to modeling measurement error, as pointed out by \cite{bartlett_bayesian_2018} and  \cite{richardson_bayesian_1993}. One such benefit is improved accuracy of estimates through better estimation of the underlying predictor function \citep{sarkar2014bayesian}. Bayesian linear regression models that allow for measurement error can be easily implemented using the Stan framework \citep{Stan_Development_Team, stan2025measurement}.
There are also more complex approaches in both the Bayesian and frequentist frameworks that rely on splines to handle measurement error in more flexible nonparametric regression models \citep{sarkar2014bayesian, jiang2023}. 






\subsubsection{Tree-based methods}

The decision tree is one of the most widely used machine learning methods. Its most popular implementation, the Classification and Regression Tree (CART) algorithm, was introduced by \cite{breiman_classification_1998}. Decision trees are built by recursively splitting the data space into rectangular partitions. Although decision trees work well in numerous scenarios, they can often become too large and start overfitting the training data. To avoid this, ensemble tree methods, such as random forests or boosting models, are used. A random forest fits trees in parallel using a bootstrapping technique to reduce overall model variance \citep{breiman_random_2001}, while boosting fits sequential trees to residuals in order to iteratively improve the model and reduce bias \citep{freund1997decision}. Decision trees and their extensions are also tolerant of missing data, interpretable, robust against collinearity, and have built-in feature selection.


Bayesian additive regression trees (BART) are a Bayesian nonparametric machine learning model first introduced by \citet{chipman_bart_2010}. Since its introduction, a variety of methods that build on the BART framework have been proposed, including approaches that favor increased model sparsity \citep{linero2018bayesian}, improve handling of categorical predictors \citep{deshpande2025flexbart}, and allow for spatially correlated input features \citep{wooten2025location}. 
For additional background on the BART model and its variants, we recommend the recent review, \citet{hill_bayesian_2020}. Notably, BART models can capture both discontinuous functions, due to the discrete nature of the splits, as well as smooth relations, due to the averaging over many weak learners.

Recent work \citep{jiang2021addressing} has demonstrated that measurement error on the inputs to random forest models can distort predictive performance and  variable importance. Although there are many variants of random forest and BART, to the best of our knowledge, there are no approaches that allow for measurement error or uncertainty in the predictors. Motivated by this gap, we propose a new Bayesian method for tree-based predictive modeling with noisy inputs.

\section{Methods}\label{sec:methods}
We now describe our proposed measurement error BART (meBART) model. We leverage the Bayesian framework to seamlessly account for uncertainty in the predictor variables used as input to the BART ensemble of trees. Below is a detailed description of our model formulation and posterior sampling approach.

\subsection{Likelihood}

Suppose there exists an i.i.d.\ sequence of random variables $\{( \boldsymbol{x}_i, y_i): i = 1,\ldots n\}$. 
Here, $\boldsymbol{x}_i$ is a $p$-vector and $y_i$ is a scalar. 
When considering the full sample, $\boldsymbol{y}$ is an $n \times 1$ column-vector, and $\boldsymbol{X}$ is an $n \times p$ matrix.
Instead of measuring $\boldsymbol{x}_i$ directly, we can only definitively measure $\boldsymbol{x}_i^*$, a noisy realization of the true $\boldsymbol{x}_i$. 
Thus, $(\boldsymbol{x}_i^*, y_i)$ are observed variables, and $\boldsymbol{x}_i$ will be considered an unknown latent variable that must be estimated. 
We use $\theta$ to denote the collection of all model parameters, which includes $\boldsymbol{x}_i$. 
We can factorize the joint likelihood for $y_i$ and $\boldsymbol{x}_i^*$ as:

\begin{equation}
p(y_i, \boldsymbol{x}_i^*|\theta) = p(y_i | \theta) \times p(\boldsymbol{x}_i^*|\theta).
\end{equation}

We can make this simplification as $\boldsymbol{x}^*_i$ and $y_i$ are conditionally independent given $\boldsymbol{x}_i$.
For the likelihood of $Y$, we build on the BART model \citep{chipman_bart_2010}. Under this approach, the expected value of $Y$ given $X$ is modeled as the sum of outputs from a set of $m$ decision trees: 

\begin{equation}
f(\boldsymbol{x}_i) = \sum\limits_{h=1}^m g(\boldsymbol{x}_i; T_h, M_h),
\end{equation}

where $T_h$ denotes the set of parameters and decision rules that govern the structure of tree $h$, and $M_h = \{\mu_{ht}: t=1, \ldots, b_h\}$ denotes the $b_h$ terminal leaf values for tree $T_h$. 
To obtain the output prediction from each tree, we follow the sequence of splits encoded by the decision rules in the tree structure until a leaf node is reached. The $\mu_{ht}$ parameter associated with that leaf node is then used as the prediction value. To achieve the final prediction for the ensemble, we sum these predictions across the $m$ trees.
By assuming i.i.d.\ errors distributed as $\mathcal{N}(0, \sigma^2$), we obtain the likelihood:


\begin{equation}
    y_i| \boldsymbol{x}_i, \mathcal{T}, \mathcal{M}, \sigma \sim \mathcal{N} \bigg( \sum\limits_{h=1}^m g(\boldsymbol{x}_i; T_h, M_h), \sigma^2 \bigg). \label{eq:likelihood}
\end{equation}

 For notational convenience, we define the collections of the tree parameters as $\mathcal{T} \defeq \{T_h:h=1, \ldots, m\}$ and $\mathcal{M} \defeq \{M_h:h=1, \ldots, m\}$.
Also as a matter of convenience, we adopt the standard assumption that the range of $\boldsymbol{y}$ has been rescaled to $[-0.5, 0.5]$ \citep{chipman_bart_2010}.






In our proposed modeling framework, we allow the predictors to be measured with uncertainty, so we also need to specify a likelihood for $\boldsymbol{x}_i^*$. We assume that the measurement error corresponds to white noise, so that the observed values are centered at the ground truth. We can then write the likelihood for $\boldsymbol{x}_i^*$ as:

%
%



\begin{equation}
    \boldsymbol{x}_i^*|\boldsymbol{x}_i,\boldsymbol{\Sigma}_e \sim \mathcal{N}_p(\boldsymbol{x}_i, \boldsymbol{\Sigma}_e).
\end{equation}

Here, we assume that $\boldsymbol{\Sigma}_e$ is a diagonal matrix with diagonal entries $\sigma^2_e$, so that the errors are independent of one another.
Alternatively, we could specify a structured covariance matrix $\boldsymbol{\Sigma}_e$ to allow for spatial or temporal correlation in the measurement error.
The joint likelihood can then be written as:
\begin{equation}
p(y_i, \boldsymbol{x}_i^*|\theta) = \mathcal{N} \bigg( \sum\limits_{h=1}^m g(\boldsymbol{x}_i; T_h, M_h), \sigma^2 \bigg) \times \mathcal{N}_p(\boldsymbol{x}_i, \boldsymbol{\Sigma}_e).
\end{equation}
\subsection{Priors}

BART uses a boosting strategy, where each new tree explains a small amount of the variance in the residuals not explained by the other trees in the ensemble. Thus, regularization priors are put on $\mathcal{T}$ and $\mathcal{M}$ to encourage each tree to be a weak learner. 
Some assumptions regarding independence of the parameters under the prior reduce the prior complexity such that: \\ \vskip-0.8cm

\begin{align}
    p(\mathcal{T}, \mathcal{M}, \sigma) \equiv p((T_1, M_1), \ldots, (T_m, M_m), \sigma) &= \bigg[ \prod_{h=1}^m p(T_h, M_h) \bigg] p(\sigma) \label{eq:joint_prior1} \\
    &= \bigg[ \prod_{h=1}^m p(M_h|T_h) p(T_h) \bigg] p(\sigma) \label{eq:joint_prior2} \\
    &= \Bigg[ \prod_{h=1}^m \bigg[\prod_{t=1}^{b_h} p(\mu_{ht}|T_h) \bigg] p(T_h) \Bigg] p(\sigma) \label{eq:joint_prior3}.
\end{align}
Thus, by equation \eqref{eq:joint_prior1}, all trees are assumed to be independent of one another under the prior, and also independent of the error variance term $\sigma$. Equation \eqref{eq:joint_prior2} shows that $M_h$ is dependent on $T_h$, which makes sense as the tree structure defines how many terminal node values are present in $M_h$, but the tree structure $T_h$ does not depend on the leaf node parameters $M_h$. Finally, equation \eqref{eq:joint_prior3} shows that the terminal node values are assumed to be independent of each other conditional on the tree, where $\mu_{ht}$ is the terminal leaf value for leaf $t$ of tree $h$. 

Now we can define the priors for the BART model parameters. The leaf means are given a normal prior, $\mu_{ht} \sim \mathcal{N}(0, \sigma_\mu^2)$, where $\sigma_\mu=0.5/(k\sqrt{m})$. Here, $k$ is a scaling factor with a typical default value of $k=2$ \citep{chipman_bart_2010}. The priors on the trees are defined by the probability of choosing a variable to split on, and, given this selection, where in the variable's support to split. Uniform priors are usually used here. To encourage shallow trees, the probability of terminating a leaf node at depth $d$ is defined as $\alpha(1+d)^{-\beta}$, where $\alpha\in (0,1)$ and $\beta\in (0, \infty)$. Finally, the variance $\sigma^2$ is given an inverse chi-square prior, $\sigma^2 \sim \nu \lambda / \chi_\nu^2$. $\nu$ and $\lambda$ are chosen such that the prior distribution assigns strong probability to the range $(0, \hat{\sigma}^2)$, where $\hat{\sigma}^2$  is either the sample variance of $\boldsymbol{y}$ or the residual variance from a least squares linear regression fit.

Under our proposed meBART approach, we treat the unobserved true predictor values as parameters to be estimated. 
We specify a normal prior on $\boldsymbol{x}_i$, the parameter that represents the unobserved true vector of predictor values for observation $i$:
\begin{equation}
\boldsymbol{x}_i \sim \mathcal{N}_p(\boldsymbol{\mu}_x, \boldsymbol{\Sigma}_x).
\end{equation}
It is possible to place a prior on $\sigma^2_e$, but in the current work we assume that it is known to ensure parameter identifiability.

\subsection{Posterior inference}




We  derive the full posterior by Bayes' rule:
\begin{align}
\text{posterior} &\propto \text{likelihood} \times \text {prior} \notag \\
p(\mathcal{T}, \mathcal{M}, \sigma^2, \boldsymbol{X}| \boldsymbol{y}, \boldsymbol{X}^*) &\propto p(\boldsymbol{y}, \boldsymbol{X}^* | \mathcal{T}, \mathcal{M}, \sigma^2, \boldsymbol{X}) \times p(\mathcal{T}) p(\mathcal{M}|\mathcal{T}) p(\sigma^2) p(\boldsymbol{X}).
\end{align}
Since this posterior is intractable, we rely on Markov chain Monte Carlo (MCMC) sampling to perform posterior inference. Our overall sampling approach is a Gibbs sampler, where we alternate between updating the BART model parameters and updating the latent parameters for the true predictor values.

Following \cite{chipman_bart_2010}, a Gibbs sampler can be applied to sample each set of BART model parameters. 
The posterior full conditional $p(\sigma|\mathcal{T}, \mathcal{M}, \boldsymbol{X}, \boldsymbol{y})$ is a standard inverse gamma distribution. In order to derive a simple form for the full conditional of the tree structures, let $T_{(h)}$ be defined as the collection of all trees except for $T_h$, and let $M_{(h)}$ be defined similarly. For the posterior full conditionals of $T_h$ and $M_h$, we make use of the fact that BART is a generalized additive model. Thus, we know that the full conditionals for each $T_h$ and $M_h$ are only dependent on $T_{(h)}$, $M_{(h)}$, $\boldsymbol{X}$, and $\boldsymbol{y}$ through the residual, $R_h=\boldsymbol{y}-\sum\limits_{k \neq h}g_k(\boldsymbol{X})$. This results in the posterior conditional $p(M_h|T_h, R_h, \sigma)$ being a normal distribution. The tree structure $p(T_h|R_h, \sigma)$ can be updated using a Metropolis-Hastings within Gibbs step.

The full conditional for the latent true values $\boldsymbol{x}_i$ is proportional to the terms in the joint posterior that involve $\boldsymbol{x}_i$:
\begin{align}
p(\boldsymbol{x}_i|\mathcal{T}, \mathcal{M}, \sigma^2, y_i, \boldsymbol{x}^*_i) &\propto p(y_i, \boldsymbol{x}^*_i | \mathcal{T}, \mathcal{M}, \sigma^2, \boldsymbol{x}_i) \times  p(\boldsymbol{x}_i) \\
&\propto \mathcal{N}(y_i ; f(\boldsymbol{x}_i|\mathcal{T}, \mathcal{M}), \sigma^2) \times \mathcal{N}_p(\boldsymbol{x}_i^*; \boldsymbol{x}_i, \boldsymbol{\Sigma}_e) \times \mathcal{N}_p(\boldsymbol{x}_i; \boldsymbol{\mu}_x, \boldsymbol{\Sigma}_x).
\end{align}

Since we can't express this distribution in closed form, we rely on a Metropolis-within-Gibbs step to sample from it. 
We propose a new $\boldsymbol{x}_i'$ based on a proposal distribution $q(\boldsymbol{x}_i'|\boldsymbol{x}_i)$. Our acceptance ratio is

\begin{equation}
\alpha = \frac{\pi(\boldsymbol{x}_i') \times q(\boldsymbol{x}_i'|\boldsymbol{x}_i)} {\pi(\boldsymbol{x}_i) \times q(\boldsymbol{x}_i|\boldsymbol{x}_i')},
\end{equation}

where $\pi(\boldsymbol{x}_i) \defeq p(\boldsymbol{x}_{i}|\mathcal{T}, \mathcal{M}, \sigma^2, y_i, \boldsymbol{x}^*_i)$. As usual, we accept the new value $\boldsymbol{x}_i'$ with probability $p = \min(1, \alpha)$. For the proposal distribution $q(\boldsymbol{x}_i'|\boldsymbol{x}_i)$, we use a random walk Metropolis-Hastings, where $q(\boldsymbol{x}_i'|\boldsymbol{x}_i) = \mathcal{N}_p(\boldsymbol{x}_i, \boldsymbol{\Sigma}_e)$. The complete Gibbs algorithm for meBART is shown in Algorithm \ref{alg:meBART}. We also detail an extension to meBART that allows for modeling binary outcomes in Algorithm \ref{alg:meBART-probit} (Appendix \hyperref[sec:Appendix_A]{A}).


\begin{algorithm}[ht]

\caption{Proposed measurement error BART (meBART) MCMC algorithm. Standard BART utilizes the same algorithm except for steps 9-11, colored blue.}
\label{alg:meBART}
\begin{algorithmic}[1]
    \While{TRUE} \Comment{One single MCMC iteration}
        \For{$h=1:m$}  \Comment{Loop over $m$ trees}
            \State Sample $T_h|R_h, \sigma^2$  \Comment{Metropolis-Hastings}
            \For{$t=1:b_h$} \Comment{Loop over $b_h$ terminal leaf nodes for tree $h$}
                \State Sample $\mu_{ht}|T_h, r_{ht}, \sigma^2$ \Comment{Normal-normal conjugate prior}
            \EndFor
        \EndFor
        \State Sample $\sigma^2|\mathcal{T}, \mathcal{M}, \boldsymbol{X}, \boldsymbol{y}$ \Comment{Normal-inverse gamma conjugate prior}
        \color{my_blue}
        \For{$i=1:n$} \Comment{Loop over $n$ observations}
            \State Sample $\boldsymbol{x}_i|\mathcal{T}, \mathcal{M}, y_i, \sigma, \boldsymbol{x}_i^*$ \Comment{New Metropolis-Hastings step}
        \EndFor
        \color{black}
    \EndWhile
\end{algorithmic}
\end{algorithm}





\section{Simulation studies}\label{sec:results}

To assess the performance of meBART, we construct multiple simulation scenarios with various synthetic data generation schemes. We start with one-dimensional data, which allows for easier visualization of the models, and then consider more complex multivariate simulation designs.

\subsection{One-dimensional synthetic data}

We begin by generating one-dimensional synthetic data, which is used as input to one of four nonlinear regression functions:

\begin{itemize}
    \item A modified indicator function ("indicator"): $f_1(x) = \begin{cases}
        -1 & \text{if } x < 0 \\
        1 & \text{if } x \ge 0
    \end{cases}$
    \item The sine function ("sin"): $f_2(x)= \sin(2\pi x)$
    \item Cosine function with discontinuity ("combo"): $f_3(x) = \begin{cases}
        \cos(\pi x) & \text{if } x < 0 \\
        -\cos(\pi x) & \text{if } x \ge 0
    \end{cases}$
    \item Step function ("step"): $f_4(x) = \begin{cases}
3 & \text{if } |x| > 5/8 \\
2 & \text{if } 3/8 < |x| \le 5/8 \\
1 & \text{if } 1/8 < |x| \le 3/8 \\
0 & \text{if } |x| \le 1/8
\end{cases}$
\end{itemize}

We start with the indicator function $f_1(x)$ because it can be thought of as a simple decision tree with one layer. The sine function $f_2(x)$ tests smooth and continuous functions, and the cosine function with discontinuity $f_3(x)$ combines the previous ideas. Finally, the step function $f_4(x)$ simulates a deeper decision tree with multiple splits.  

$X$ is generated such that $x_i \sim \text{ i.i.d. } \mathcal{N}(0, 0.3^2)$. We then add Gaussian measurement error to generate the measured data $x_i^*=x_i+e_i$, where $e_i \sim \mathcal{N}(0, 0.1^2)$. We add an identically distributed Gaussian error to the output variable, such that $y_i=f(x_i)+\varepsilon_i$ and $\varepsilon_i\sim \mathcal{N}(0, 0.1^2)$. 100 of the 200 total data points are used to train each model, and the remaining 100 data points are reserved for testing. 

We compare our proposed method, meBART, to four alternative approaches: Bayesian linear regression (BLR), Bayesian linear regression with measurement error (BLRME), random forest, and BART. All simulations were programmed in R \citep{R}. BLR was fit using the bayesreg package \citep{bayesreg}, BLRME using Stan \citep{Stan_Development_Team} and the CmdStanR package \citep{cmdstanr}, random forest using the randomForest package \citep{randomForest}, and BART using the BART package \citep{sparapani_nonparametric_2021}.
We compare the five methods using six metrics to assess accuracy of prediction, recovery of the true function $f(x)$, and estimation of the standard deviation $\sigma$. 

We consider two metrics of prediction error: 
prediction mean squared error (MSE) calculated using the noisy $X^*$, and MSE using the true $X$ values for the independent test dataset.
The noisy MSE calculation is a standard measure of predictive performance on unseen data, while the true $X$ MSE calculations measure how well the method would perform on a new dataset with limited noise. MSE using the true $X$ also serves as a proxy of how well the method is estimating the true underlying function. To more directly assess accuracy in estimation of the true underlying function, we compute the integrated square error (ISE), where $\text{ISE} = \int (f_{true}(x) - f_{est}(x))^2 \, dx$. MSE and ISE provide complementary information: MSE using the true values for the test dataset gives more weight to the denser areas of the $X$ domain, while ISE, which is calculated over the entire support of $X$, equally weights all areas of the $X$ domain. We calculate the root mean squared error in estimation of the true predictor values ($X$ RMSE) to show how well the measurement error models are learning the true values of $X$. Finally, we provide the continuous ranked probability score (CRPS) for both the output function posterior estimates and the posterior draws of $\sigma$. CRPS is a distributional generalization of mean absolute error (MAE), and provides an additional measure of the precision and accuracy of the estimated functions \citep{gneiting2007strictly}. 

\begin{figure}[ht]
\centering
     \includegraphics[width=\textwidth]{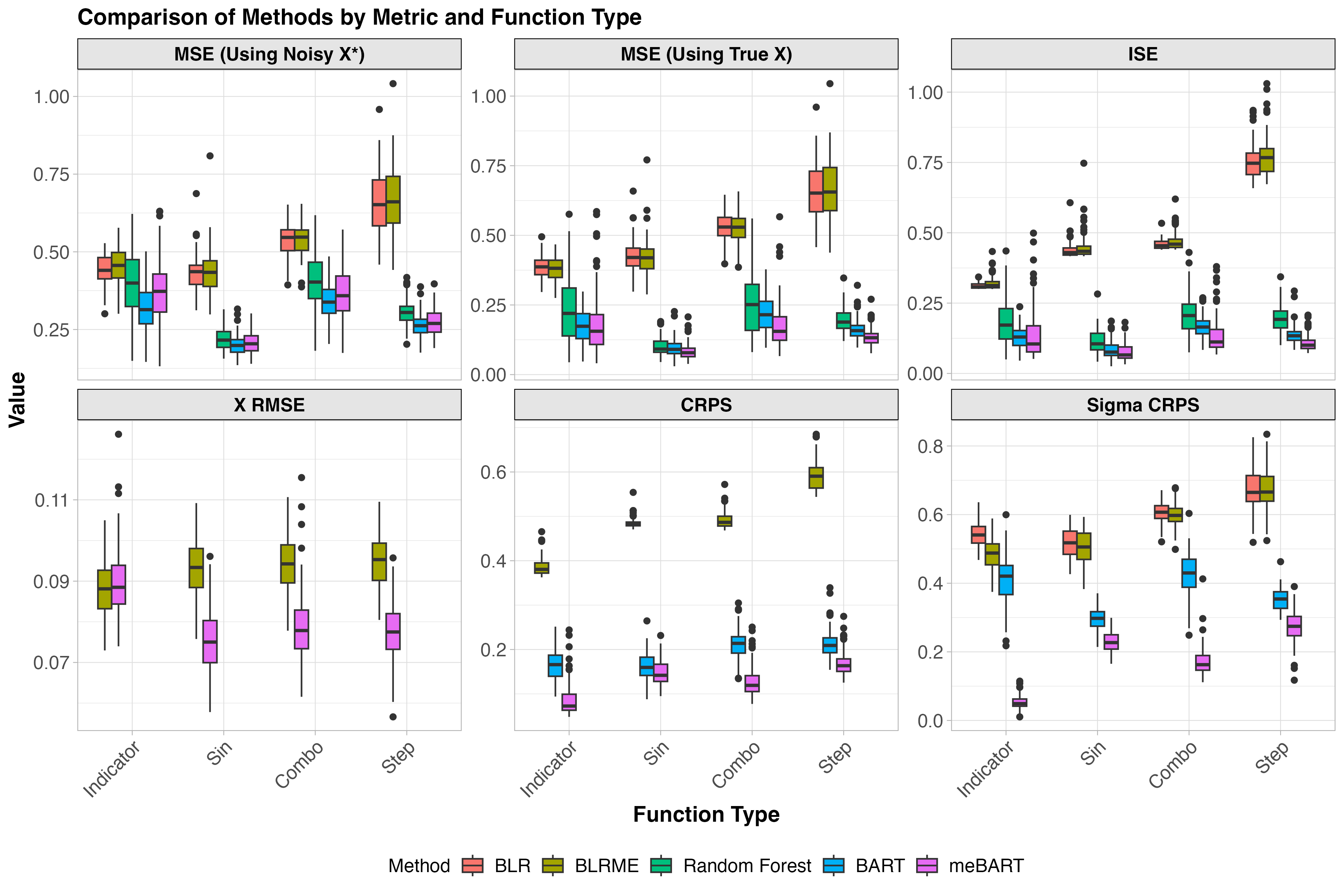}
     \caption{One-dimensional input data simulation results summarized over 100 simulated datasets. }
     \label{fig:1D-simulation-results}
 \end{figure}

In Figure \ref{fig:1D-simulation-results}, we illustrate the performance of all five methods summarized across 100 simulated data sets. BART, meBART, and random forest perform consistently better on traditional MSE than the linear methods tested. This is not very surprising as the functions being tested are nonlinear. Although still competitive, meBART's performance is somewhat more variable on the combo and indicator function datasets. 

The MSE using true $X$ values yields similar patterns, but with generally lower MSE values. meBART yielded the lowest median values for each of the four function types. ISE performance is generally consistent with that of MSE using the true $X$ values, which is expected as they have similar calculations. The major difference between the two metrics is that MSE is calculated on an independent test dataset (drawn from the same $X$ prior), while ISE is calculated over the entire domain of the function. meBART also yields the lowest median ISE values for each of the four function types.

meBART and BLRME, the only two methods that estimate the latent $X$ values, both provide estimates of $X$ that are closer to the true values than the measured values $X^*$, while meBART has even closer estimates than BLRME. 
Finally, the CRPS values provide a sense of how well the posterior distributions are modeling the true underlying functions $f_1, \ldots, f_4$ and $\sigma$. CRPS rewards both precise and accurate estimates. In both cases, meBART exhibits excellent performance with the lowest CRPS values, followed by BART.

\begin{figure}[ht]
\centering
     \includegraphics[width=0.75\textwidth]{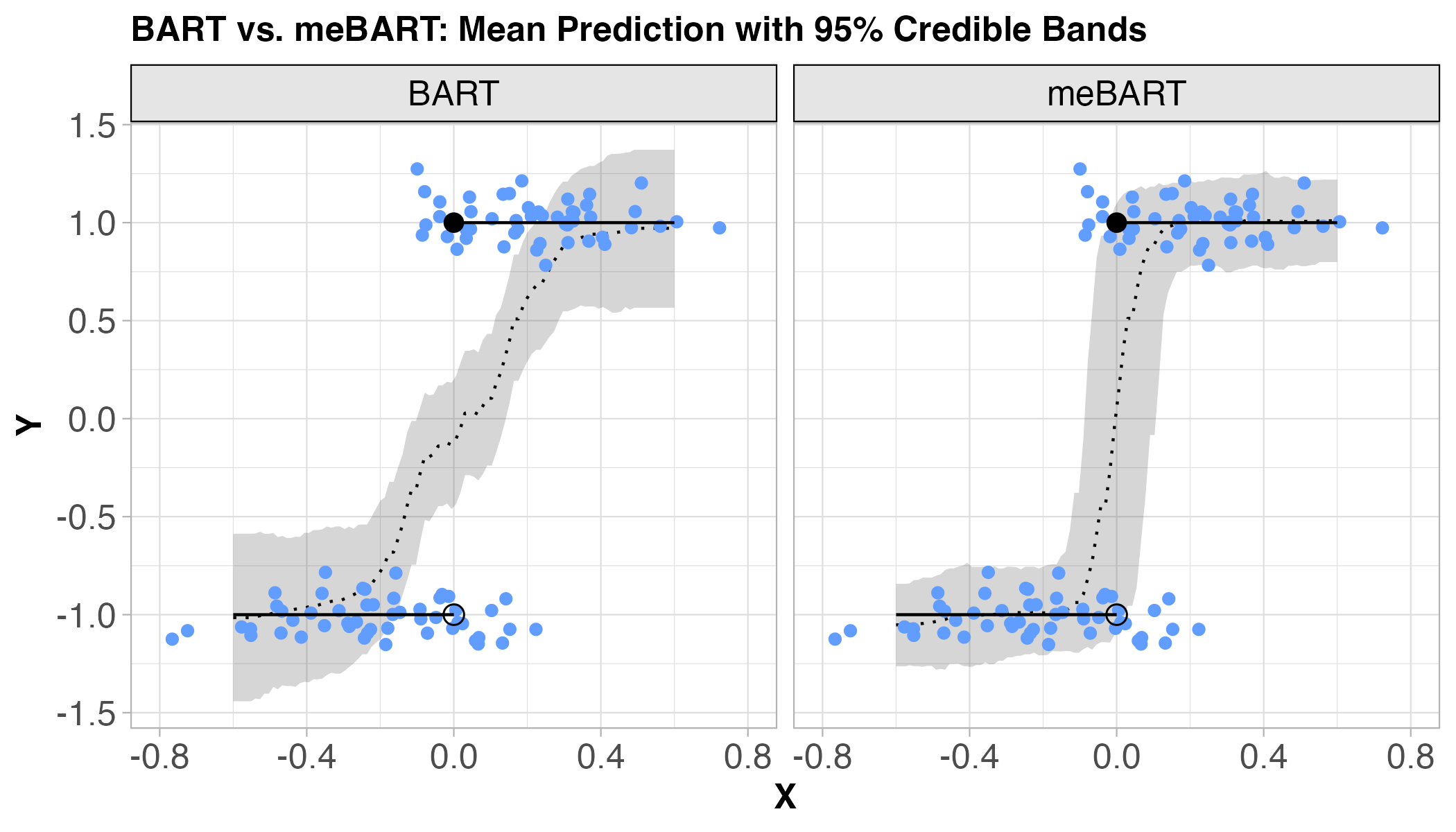}
     \caption{Comparison of BART and meBART function estimation and 95\% credible intervals for one simulated data set from the indicator function setting. The true underlying function is plotted as a solid black line and the data points with measurement error added are shown as blue dots. The dotted black lines and gray regions represent the point-wise mean and point-wise 95\% posterior credible interval. 
     }
     \label{fig:CIs}
 \end{figure}

To better visualize the benefits of meBART, we provide a sample plot comparing the function estimation and credible intervals obtained from BART and meBART in Figure \ref{fig:CIs}. meBART provides a much smoother estimated function, with a sharper transition that more accurately captures the underlying true discontinuity. 
This pattern is also observed when we use a 2-dimensional analog to the indicator function, $f(x_1, x_2) = \mathbbm{1}(x_1>0) + 2\times\mathbbm{1}(x_2>0)$, as shown in Figure \ref{fig:2d-contours} (Appendix \hyperref[sec:Appendix_C]{C}). 
meBART also provides much better empirical coverage across the discontinuity, which is where most of the data points lie.

\begin{figure}[ht]
\centering
     \includegraphics[width=0.6\textwidth]{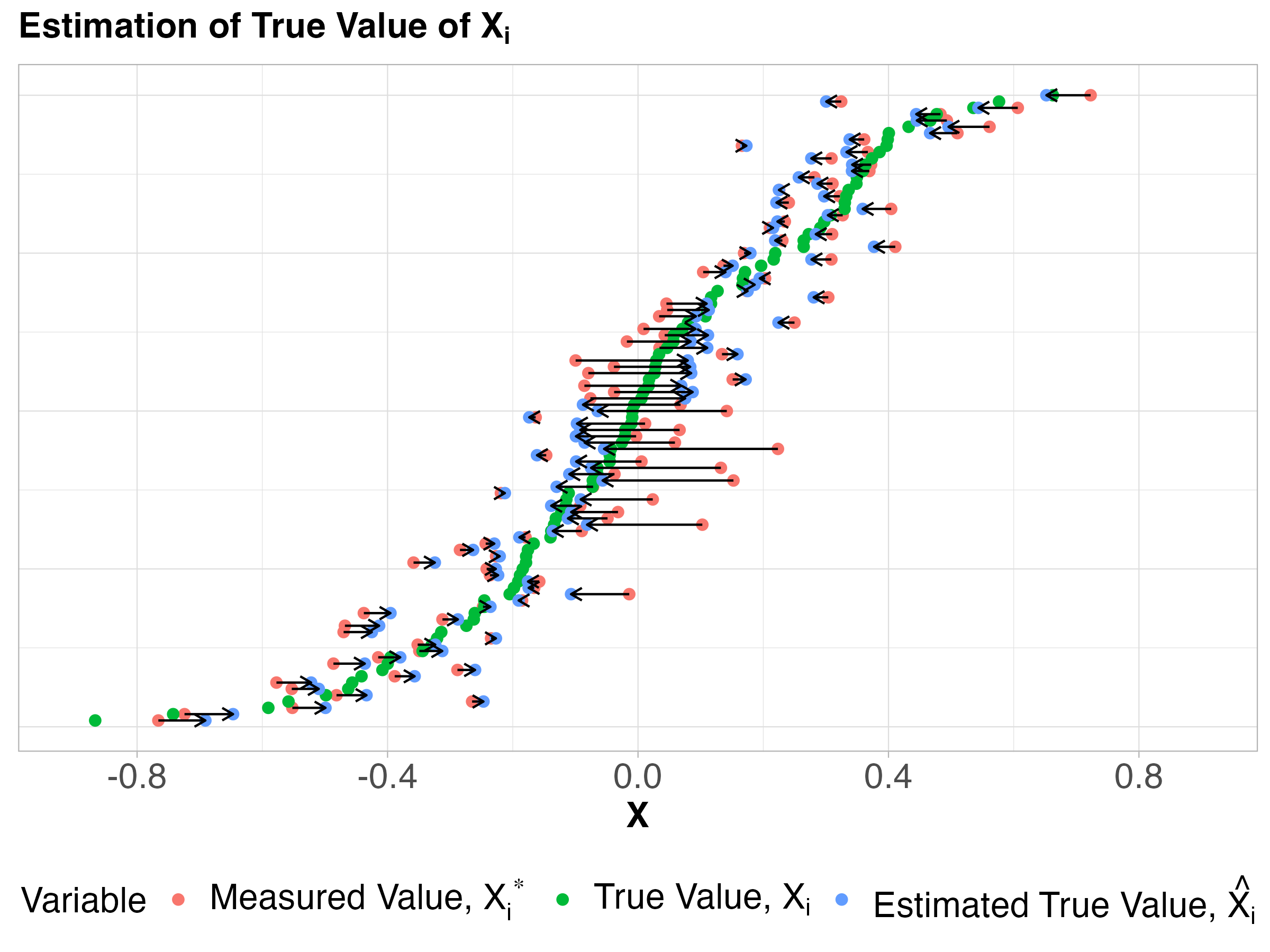}
     \caption{The true $x_i$ variables, the observed noisy $x_i^*$, and the posterior estimates of $x_i$, $\hat{x}_i$. Arrows point from $x_i^*$ to $\hat{x}_i$ and represent our updated beliefs about the true value of $x_i$.}
     \label{fig:x_draws}
 \end{figure}

 Finally, we plot the true $\boldsymbol{x}$ values, the observed noisy $\boldsymbol{x}^*$, and the posterior estimates $\hat{\boldsymbol{x}}$ in Figure \ref{fig:x_draws}. The arrows in the figure point from $x_i^*$ to $\hat{x}_i$, and represent our updated beliefs about the true value of $x_i$. Most of the estimates $\hat{x}_i$ move correctly in the direction of their respective true value $x_i$. Larger differences between $x_i^*$ and $\hat{x}_i$ are concentrated around $0$, which is the point of discontinuity for the indicator function used in this test dataset. This makes sense, as points far away from the discontinuity, where the underlying true function is flat, don't have any signal or variation in $\boldsymbol{y}$ that would guide an updated estimate of $x_i$.

\subsection{Multi-dimensional synthetic data}

To evaluate the performance of meBART on multivariate input data, we make use of the Friedman function \citep{friedman_multivariate_1991}, a common benchmark function used in machine learning:

\begin{equation}\label{eq:friedman}
    f(x) = \sin(\pi x_1 x_2) + 2(x_3-0.5)^2 + x_4 + 0.5x_5.
\end{equation}

We generate $X$ such that $\boldsymbol{x_i} \sim \text{ i.i.d. } \mathcal{N}(0.5, 0.3^2\times \mathbb{I}_5)$ and add independent measurement errors sampled from  $\mathcal{N}(0, \sigma_e^2 \times \mathbb{I}_5)$ to generate $\boldsymbol{X}^*$. In this set of simulation scenarios, we use four different values of $\sigma_e \in \{0.05, 0.1, 0.15, 0.2\}$. As in the univariate simulation, we compare meBART against BLR, BLRME, random forest, and BART. In the multivariate simulations, we replace the ISE performance metric with the empirical coverage probability at a 95\% credible interval, since ISE can become prohibitively expensive to compute across multiple dimensions. We also scale the $X$ RMSE values by $\sigma_e$ so that we can better gauge accuracy in estimation of $X$ for larger amounts of measurement error. For large $n$, the RMSE of $X^*$ should have an average value of $\sigma_e$.

\begin{figure}[ht]
\centering
     \includegraphics[width=\textwidth]{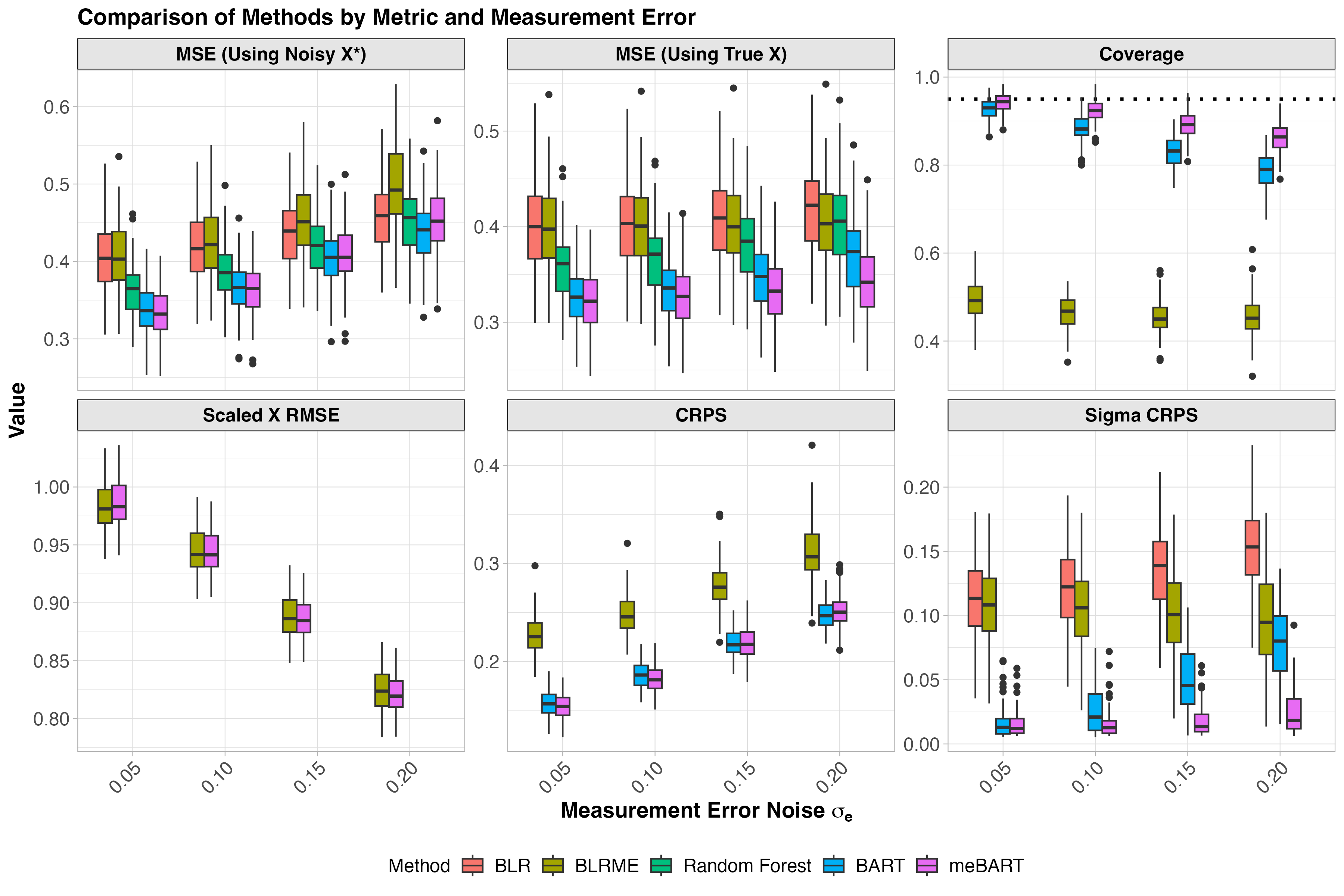}
     \caption{Multidimensional simulation setting results summarized over 100 independently generated datasets.}
     \label{fig:5D-simulation-results}
 \end{figure}

The results from these simulation settings are shown in Figure \ref{fig:5D-simulation-results}. meBART yields similar MSE to BART throughout the range of measurement error standard deviations $\sigma_e$ tested. 
meBART outperforms BART on MSE for the true $X$ values by an increasing margin as $\sigma_e$ is increased. The empirical coverage for meBART is closest to the theoretical value of 0.95 in all four settings tested. $X$ RMSE shrinks as measurement error gets larger, indicating that as the amount of measurement error increases, the capacity to learn the true underlying $X$ values improves. meBART achieves the lowest CRPS for the $\boldsymbol{y}$ estimates in three of the four settings, though BART is a close second across the multiple noise levels. Finally, meBART yields the best sigma CRPS among all of the Bayesian methods where this metric is computed.

\subsection{Convergence assessment}

\begin{figure}[ht!]
\centering
     \includegraphics[width=0.6\textwidth]{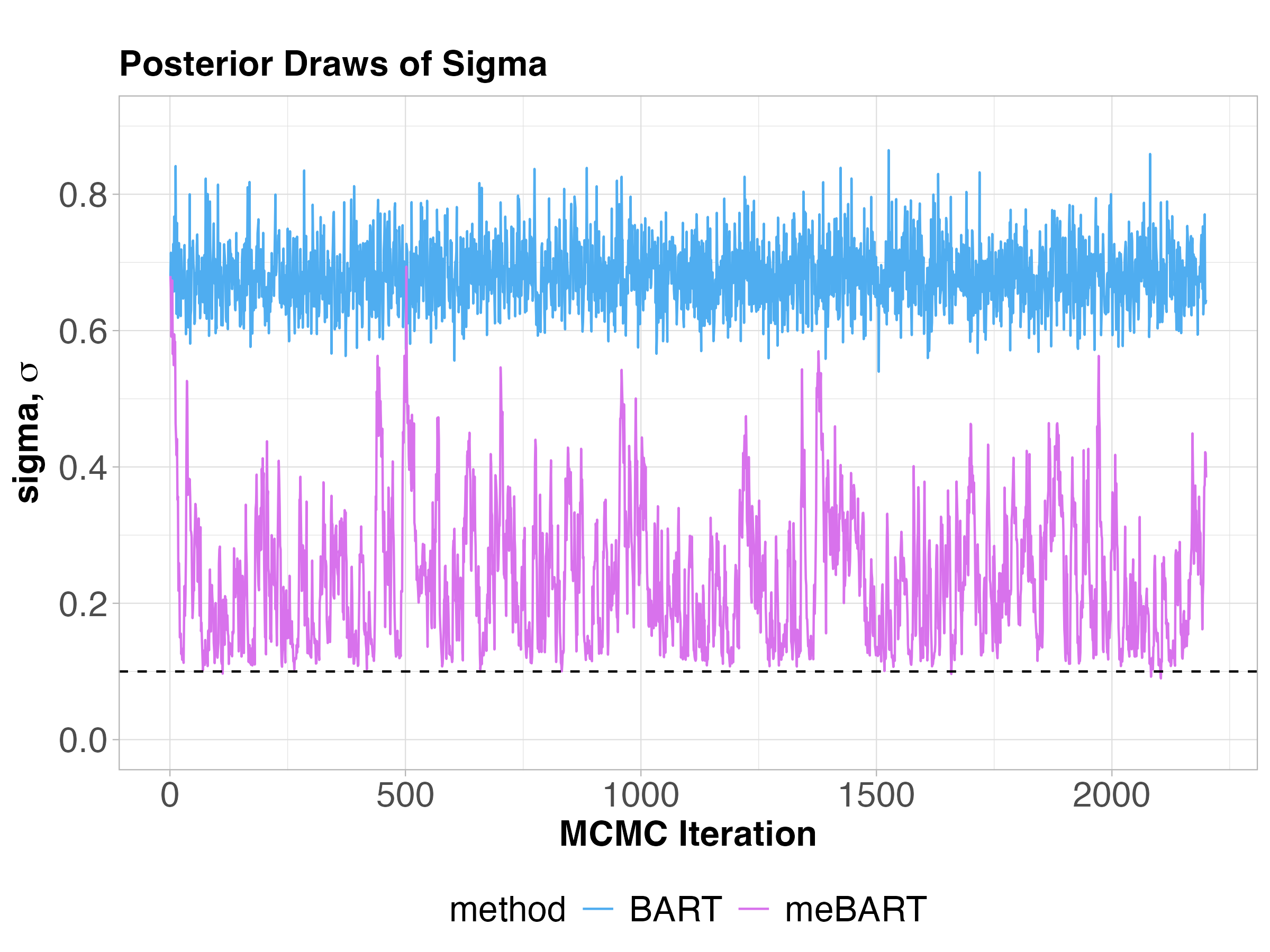}
     \caption{$\sigma$ trace plots for one simulated data set from the indicator function setting. Both BART and meBART utilized 200 MCMC iterations for burn-in. The dashed black line represents the true underlying value of $\sigma=0.1$.}
     \label{fig:sigma}
 \end{figure}

Due to the changing number of tree parameters and their lack of inherent meaning, posterior convergence of BART models is usually assessed by trace plots of the error variance $\sigma$ \citep{sparapani_nonparametric_2021}. An example $\sigma$ trace plot is provided in Figure \ref{fig:sigma}. While meBART doesn't estimate the $\sigma$ parameter perfectly, it is much closer than standard BART. It is also noticeable that the mixing of meBART is slightly poorer. 
Finally, you can tell from the left-hand side of the plot that meBART burns in very quickly. 

\section{Real data applications: microbiome diversity and self-reported activity data} \label{sec:real_data}

In this section, we consider two biomedical applications in which the goal is to predict a clinical response variable from a predictor that is known to be subject to measurement error. In both case studies, the relationship between the predictor and response is hypothesized to be nonlinear, motivating the need to use a flexible model such as meBART to characterize the predictor-outcome relationship.

\subsection{Prediction of stem cell transplant outcomes from microbiome diversity}

The human gut is populated by hundreds of species of bacteria, constituting a dynamic community that plays an important role in sustaining normal digestive function and immune activity \citep{ghosh2021structural}. A diverse gut microbiome is associated with healthy metabolism and resistance to pathogenic infection, while a disrupted or imbalanced microbiome has been associated with a variety of disorders, including obesity, inflammatory bowel disease, and  diabetes \citep{backhed2012defining}. As discussed in the Introduction (Section \ref{sec:introduction}), quantifications of microbiome diversity are subject to substantial measurement error due to sampling variability, technical noise, and the underlying dynamics of the system \citep{vogtmann2017comparison, ji2019quantifying, willis2019rarefaction}.

Here, we consider the challenge of predicting stem cell transplant outcomes from  quantifications of gut microbial diversity. 
In allogeneic stem cell transplantation, hematopoietic stem cells from a donor are infused into a patient in an effort to restore the production of normal healthy blood cells. Stem cell transplantation is a key therapy for hematologic diseases such as  leukemia and  myeloma \citep{Singh2016}. The goal is for the infused cells to engraft, or find their way to the bone marrow and begin producing new healthy blood cells. Since subjects are vulnerable to infection and anemia in the interval between transplantation and engraftment, a lower time to engraftment is considered a positive clinical outcome.

We utilized a data set originally collected by \cite{schluter2020gut} to probe the relationship between microbiome diversity and time to stem cell engraftment. 
\cite{Peled2020} demonstrated that higher microbiome diversity was associated with decreased risk of death across multiple independent cohorts of stem cell transplant recipients, confirming the results of previous studies conducted at single institutions \citep{Taur2014, Golob2017}. 
Here, we seek to predict the number of days from the bone marrow transplant until engraftment as a key clinical response variable. 
We filtered the data set to include patients who had received peripheral blood stem cells, did not receive more than one bone marrow transplant, and did not experience graft failure, resulting in a 
dataset of $n=375$ patients.  
As our predictor, we rely on
the inverse Simpson index, a standard metric of microbial alpha diversity; higher values for this index reflect a more diverse gut microbiome \citep{xia2023alpha}. 
Since the distribution of the inverse Simpson values was highly right-skewed, we applied a log transformation prior to downstream analysis.

\begin{figure}[ht]
\centering
     \includegraphics[width=0.75\textwidth]{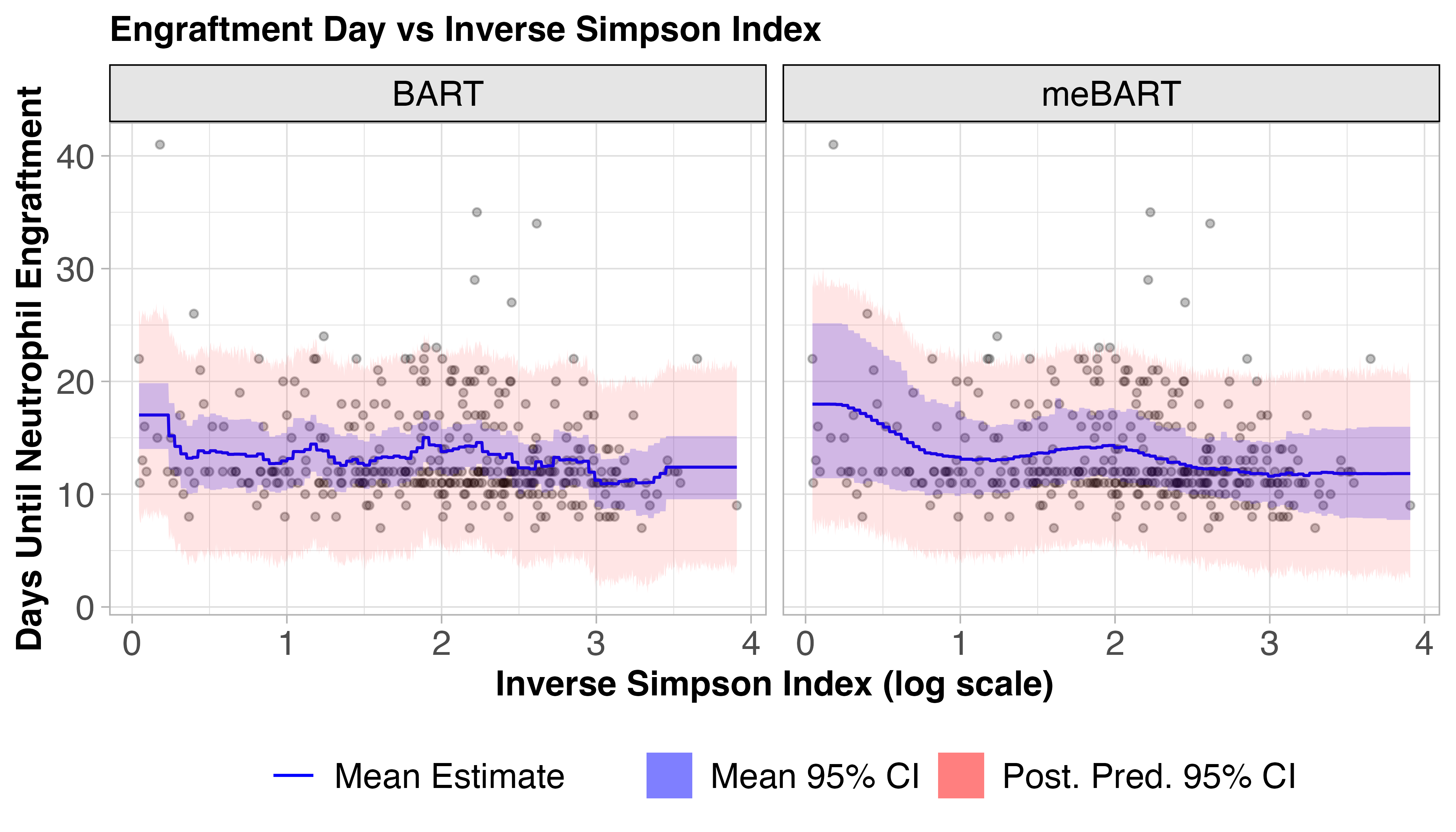}
     \caption{Results of BART and meBART for prediction of engraftment day as a function of inverse Simpson index. The mean estimates are shown as a solid blue line, with the 95\% credible interval for the mean shown as shaded blue regions and the 95\% posterior predictive credible bands shown as shaded red regions.}
     \label{fig:edayVsSimpson}
 \end{figure}

We applied BART and meBART to learn the predictor-outcome relationship. The posterior estimates of the mean, 95\% credible interval, and 95\% prediction intervals for each method are shown in Figure \ref{fig:edayVsSimpson}. 
Overall, the learned functions align with our expectation that 
 microbial diversity and time to engraftment should have a negative relationship, as a more diverse microbiome has been associated with improved response to bone marrow transplant. Both methods infer a nonlinear relationship, with a relatively sharp decrease in risk as the log of the inverse Simpson index increases from 0 to 1, followed by a relatively flat risk profile for higher diversity values. The inferred function from meBART is notably smoother and has wider posterior credible intervals, reflecting the propagation of uncertainty regarding the true, unobserved diversity encoded in the measurement-error framework. 
 
 Our results suggest that very low microbiome diversity, possibly resulting from antibiotic or chemotherapy received prior to transplant, largely drives the previously reported association between microbiome diversity and poor stem cell transplant outcomes \citep{Peled2020}. These results also motivate the design of microbiome interventions, such as prebiotics, probiotics, or specially-designed combinations of these, to "rescue" patients with highly depleted gut microbiomes at baseline \citep{khalil2024impact}.


\subsection{Relationship between self-reported activity and body mass index}
Here, we seek to characterize the relationship between physical activity levels and body mass index (BMI). To investigate this association, we utilize a dataset from a recent study on cardiometabolic disease \citep{fromentin2022microbiome}, which estimated subjects' physical activity based on the Recent Physical Activity Questionnaire (RPAQ). Self reported diet and activity estimates are known to be subject to measurement error \citep{wang2024biomarker}.
Previous large-scale studies have shown that estimates of activity from the RPAQ have only a moderate correlation with gold-standard estimates of physical activity obtained using a combination of movement sensing and heart rate monitoring \citep{golubic2014validity}. However, since activity surveys are much cheaper and easier to administer than these more intense monitoring approaches, they remain common in many biomedical studies. 

In their publication, \cite{fromentin2022microbiome} performed a simple comparison across groups, finding that healthy subjects had significantly higher reported physical activity (hours/week) than subjects with ischemic heart disease. 
 However, the benefits of exercise are likely to be nonlinear. Previous studies have shown that even minimal levels of exercise improve life expectancy compared with being inactive \citep{wen2011minimum}. The expected benefit of increased activity flattens out as the volume of activity increases, particularly for vigorous activity \citep{eijsvogels2016exercise}. 
 
\begin{figure}[ht]
\centering
     \includegraphics[width=0.75\textwidth]{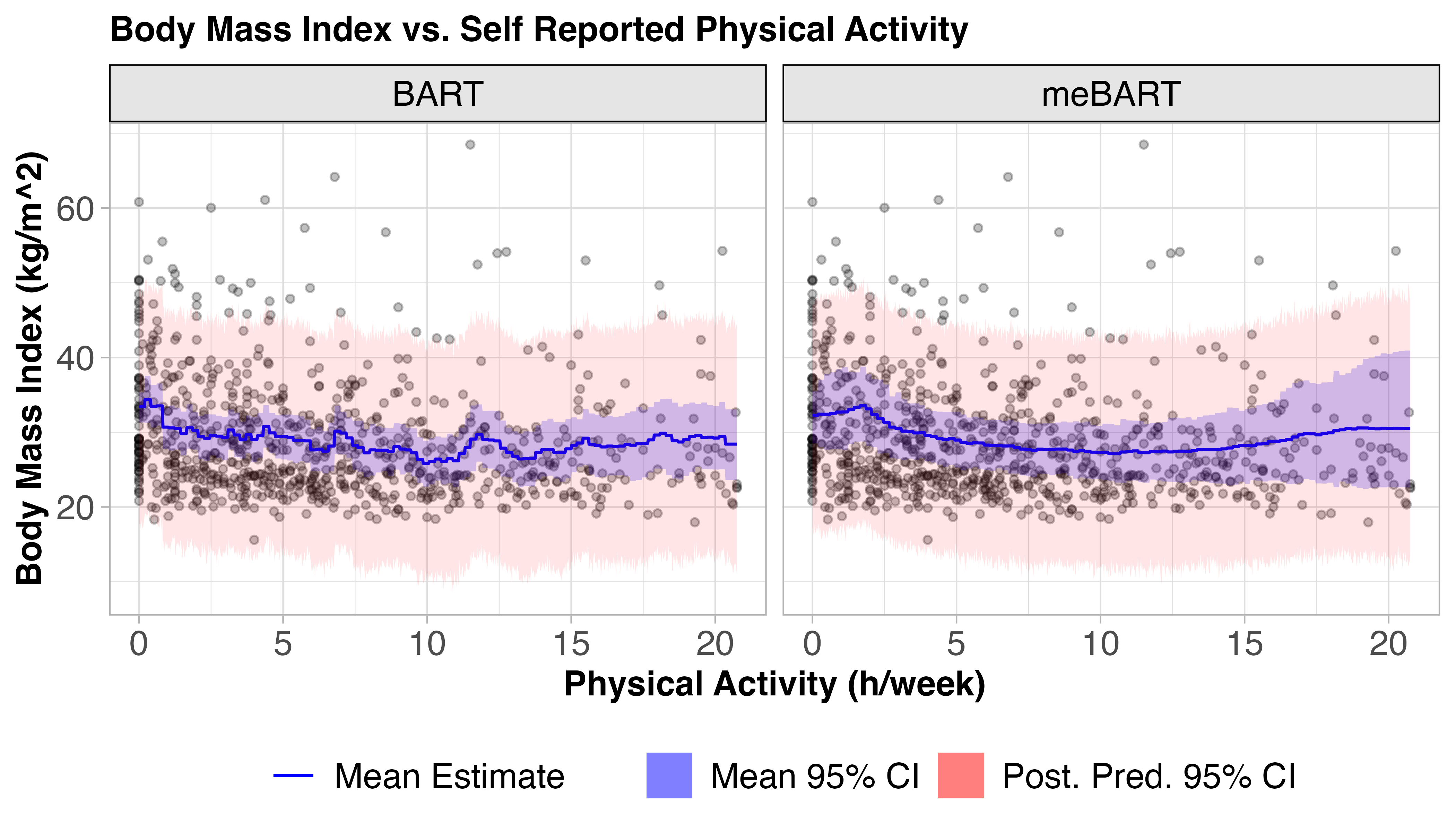}
     \caption{Results of BART and meBART for prediction of body mass index (BMI) as a function of self-reported physical activity levels. The mean estimates are shown as a solid blue line, with the 95\% credible interval for the mean shown as shaded blue regions and the 95\% posterior predictive credible bands shown as shaded red regions.}
     \label{fig:BMIvsPA}
 \end{figure}

To showcase the utility of meBART, we seek to predict body mass index (BMI) as a function of reported physical activity per week. 
The dataset obtained from \cite{fromentin2022microbiome} includes 809 subjects with reported physical activity estimates. We applied both meBART and BART to learn the predictor-outcome relation. 
We plot the estimated means obtained from each method in Figure \ref{fig:BMIvsPA}. 

meBART provides a much smoother overall function, suggesting that it is less prone to overfitting than BART. meBART also has wider credible intervals where there is less support in the data, particularly in the tail regions. 
\cite{eijsvogels2015exercise} point out that such a J-shaped relationship between exercise and BMI is likely, with moderate amounts of exercise being beneficial for cardiovascular health but with diminishing returns beyond a certain point. 
Our results support a roughly J-shaped relation between exercise levels and BMI, with  a slight increase in predicted BMI for very high exercise levels.  This could reflect additional muscle mass in subjects with high activity levels, as BMI does not differentiate between fat vs.\ non-fat mass \citep{allison2021addressing}. 
Importantly, given the observational design of the study, the inferred relationship between exercise and BMI has no causal interpretation; it is also possible that subjects with high BMI may be adopting strict exercise regimens.

To assess predictive performance, we report the average mean squared error (MSE) across 5-folds of  each case study data set in Table \ref{tab:realdata_mses}. We observe that meBART has the lowest average MSE for the case study on prediction of BMI from physical activity, and performs similarly to BART in prediction of engraftment day from microbiome diversity. BART and meBART both outperform a classic random forest.

\begin{table}[ht]
    \centering
        \caption{Average mean squared error (MSE) and standard error of 5-fold cross validated prediction.\\ \vspace{.1cm}}
    \begin{tabular}{cccccc}
    \toprule
    dataset &  RF &  BART & meBART \\
    \midrule
    Engraftment Day vs. Inverse Simpson Index & 25.16 $\pm$ 2.13 & \textbf{19.12 $\boldsymbol{\pm}$ 2.66} &  19.46 $\pm$ 2.85 \\
    Body Mass Index vs. Physical Activity &  75.46 $\pm$ 3.78  & 60.99 $\pm$ 3.49 & \textbf{59.99 $\boldsymbol{\pm}$ 3.17} \\
    \bottomrule
    \end{tabular}
    \label{tab:realdata_mses}
\end{table}

\section{Discussion}\label{sec:discussion}

Based on the results from our simulation and real data analyses, 
we observe that meBART achieves similar predictive performance to BART in terms of MSE, with improved accuracy in function estimation and uncertainty quantification for both predictions and model parameters. meBART has the added benefit of giving refined estimates of the true values of $X$.

One interesting advantage of meBART is the ability to produce much smoother function estimates than BART. As shown in Figure \ref{fig:CIs}, the estimated function mean, the 2.5\% quantile, and the 97.5\% quantile produced by meBART are all much smoother than those produced by BART. This pattern was found across all of the one-dimensional test functions. The smoothing effect makes intuitive sense if you think about the model `jiggling' the true $\boldsymbol{x}_i$ values around their initial estimate of $\boldsymbol{x}_i^*$. This results in observations occasionally drifting over the decision boundaries of the internal nodes in the trees of the BART model. This is reminiscent of soft BART (SBART), which treats the tree decision rules as probabilistic gates \citep{linero2018bayesian}. 


The meager boosts to MSE performance may come as a surprise given the improvements in other metrics as well as the visually superior function recovery seen in Figure \ref{fig:CIs}. This is likely due to how the squaring of the error punishes meBART for making more definitive estimates near the discontinuity of the indicator and combo datasets. When calculating MSE on a noisy input, like $(-0.2, 1)$, meBART would make a prediction close to $-1$, while BART makes an intermediate prediction of $0$, which leads to a smaller squared error.

This phenomenon is related to a larger struggle that measurement error models face. In the presence of noisy input variables, a na\"ive model learns a function that maps $f_{\text{na\"ive}}:X^* \to Y$, while a measurement error model tries to learn the true mapping $f_{me}:X \to Y$. However, when making predictions on additional noisy data $\boldsymbol{X}^*$, a na\"ive model may lead to better performance. Essentially, measurement error models can provide estimates of the true $X$ values and improved characterization of the relationship between $X$ and $Y$, but may not necessarily improve predictive accuracy on future noisy data.
 




The proposed meBART method has several limitations. 
The primary drawback of meBART vs. BART is that fitting the meBART model entails estimation of $n$ additional parameters, increasing the complexity of posterior sampling. 
As shown in Appendix \hyperref[sec:Appendix_B]{B},  this results in longer runtimes for meBART, especially as $n$ increases. 
We also find that meBART does not recover the true $X$ values perfectly. In the one-dimensional data simulation, we find that meBART reduces noise in $X$ by a factor of about $0.8$. However, meBART still offers better recovery of the true $X$ than the nonlinear measurement error method, BLRME, which reduces the noise in $X$ by a factor of about $0.9$ across the simulations. There are a few reasons why this could be. For one, better $X$ estimates are obtained where the support of $X$ is strongest. If there isn't enough signal in the data to learn the function, there is also not enough signal to learn the true underlying value of $X$.


There are a number of potential areas of interest for future work. In particular, we could explore other distributional assumptions for the predictors and the measurement error. 
Measurement error on a categorical predictor, known as misclassification, would be useful for many applications but would require a different likelihood for $X$ \citep{koslovsky2024unified}.
In this work, we focused on settings where the observations were i.i.d. Future work could be done to extend the method to data sets with clustered or longitudinal observations. 
It would also be of interest to consider settings with systematic or correlated measurement error. In the current work, we focused on random measurement error, in the sense that the mean of the errors is assumed to be 0.
Non-symmetrical distributions where the measurement error is biased in either a positive or negative direction could be appropriate for certain applications. 
For example, an application to voter polling data could have spatially correlated measurement error across neighboring counties. 


\section{Conclusion}\label{sec:conclusion}



Measurement error in the input variables of a prediction model can have profound effects if unaccounted for. 
Little research has been done to address measurement error in the inputs to machine learning methods. 
Here, we proposed the meBART method, which combines the flexible modeling of BART with the improved uncertainty quantification of measurement error models. 
meBART performs better than the competing methods in function recovery and posterior estimation of the true input values. 
meBART also provides more accurate coverage probabilities and more accurate estimation of noise variance parameters. 
Although meBART offers similar prediction performance on independent noisy test data when compared to other models, it offers the added benefit of learning estimated values for the noisy input variables that are closer to the ground truth.

\section*{Code availability}
To enhance computational scaling, we implement meBART as an Rcpp package and an extension of the BART package by \cite{sparapani_nonparametric_2021}. meBART is available to download at \url{https://github.com/kmccoy3/meBART}.
The GitHub repository includes instructions for installation, vignettes with basic use cases, and the code used to generate the plots contained in this manuscript.

\section*{Acknowledgements}

K.M.\ is supported by the National Science Foundation (NSF) Graduate Research Fellowship Program (GRFP) under Grant No. 1842494, and by the Ken Kennedy Institute Computational Science and Engineering Recruiting Fellowship, funded by the Energy HPC Conference. C.B.P.\ is partially supported by NIH/NHLBI R01 HL158796 and by NIH/NCI P01 CA278716 (Core B - Biostatistics). K.M.\ thanks Tyler Bagwell for reviewing and suggesting edits to this manuscript.

\newpage
\bibliographystyle{plainnat}
\bibliography{main}

\newpage
\appendix

\section{Probit classification for meBART}\label{sec:Appendix_A}

Extending meBART to qualitative outputs is relatively easy, as probit BART only requires a few minor modifications to the likelihood. Probit BART uses the likelihood function:

\begin{equation}
    p(y_i=1|\theta) =\Phi \bigg[\sum\limits_{h=1}^m g(\boldsymbol{x}_i; T_h, M_h) \bigg].
\end{equation}

Note that the error variance $\sigma$ is no longer present in the model. 
Probit BART utilizes latent variables $z_1, \ldots, z_n \sim \mathcal{N}(f(\boldsymbol{x}_i), 1)$ such that $y_i=1$ if $z_i>0$, and $y_i=0$ if $z_i \leq 0$. The full conditional is then simply $z_i|[y_i=1] \sim \max\{\mathcal{N}(f(\boldsymbol{x}_i), 1), 0\}$ and $z_i|[y_i=0] \sim \min\{\mathcal{N}(f(\boldsymbol{x}_i), 1), 0\}$. We can then define our posterior sampling algorithm as follows:

\begin{algorithm}[ht]
\caption{Measurement Error BART (meBART) MCMC with probit classification. Regular probit BART follows the same except for steps 11-13, colored blue.}
\label{alg:meBART-probit}
\begin{algorithmic}[1]
    \While{TRUE} \Comment{One single MCMC iteration}
    \For{$i=1:n$} \Comment{Loop over $n$ observations}
            \State Sample $z_i|y_i, \mathcal{T}, \mathcal{M}, \boldsymbol{X}$ \Comment{Sample latent variable using truncated normal}
        \EndFor
        \For{$h=1:m$}  \Comment{Loop over $m$ trees}
            \State Sample $T_h|R_h$  \Comment{Metropolis-Hastings}
            \For{$t=1:b_h$} \Comment{Loop over $b_h$ terminal leaf nodes for tree $h$}
                \State Sample $\mu_{ht}|T_h, r_{ht}$ \Comment{Normal-normal conjugate prior}
            \EndFor
        \EndFor
        \color{my_blue}
        \For{$i=1:n$} \Comment{Loop over $n$ observations}
            \State Sample $\boldsymbol{x}_i|\mathcal{T}, \mathcal{M}, \boldsymbol{z}, \boldsymbol{x}_i^*$ \Comment{New Metropolis-Hastings step}
        \EndFor
        \color{black}
    \EndWhile
\end{algorithmic}
\end{algorithm}

\section{Runtime analysis}\label{sec:Appendix_B}

\begin{figure}[ht]
\centering
     \includegraphics[width=0.6\textwidth]{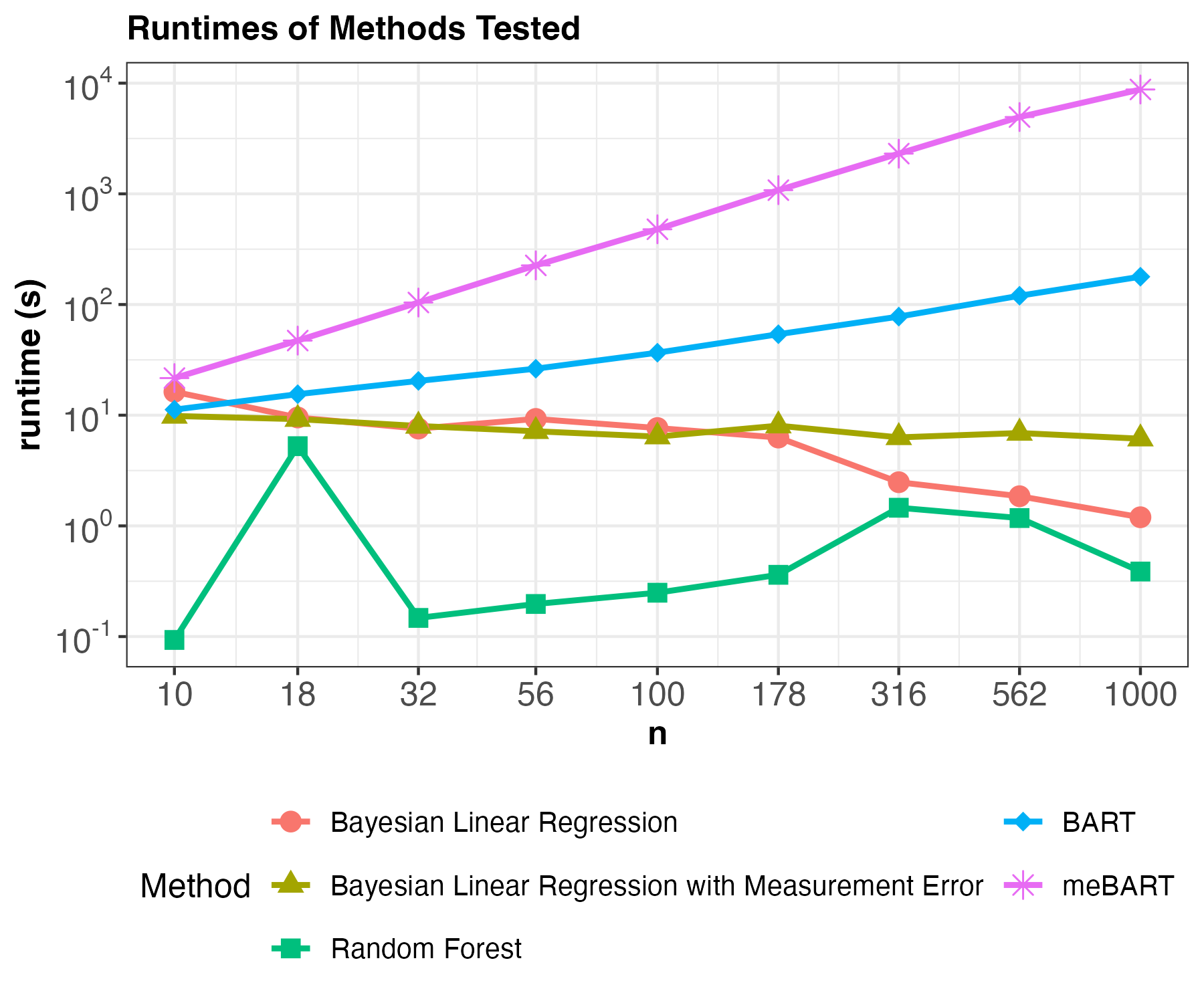}
     \caption{Runtimes of various methods tested.}
     \label{fig:runtimes}
 \end{figure}

 \section{2-Dimensional indicator function}\label{sec:Appendix_C}

\begin{figure}[ht]
\centering
     \includegraphics[width=0.9\textwidth]{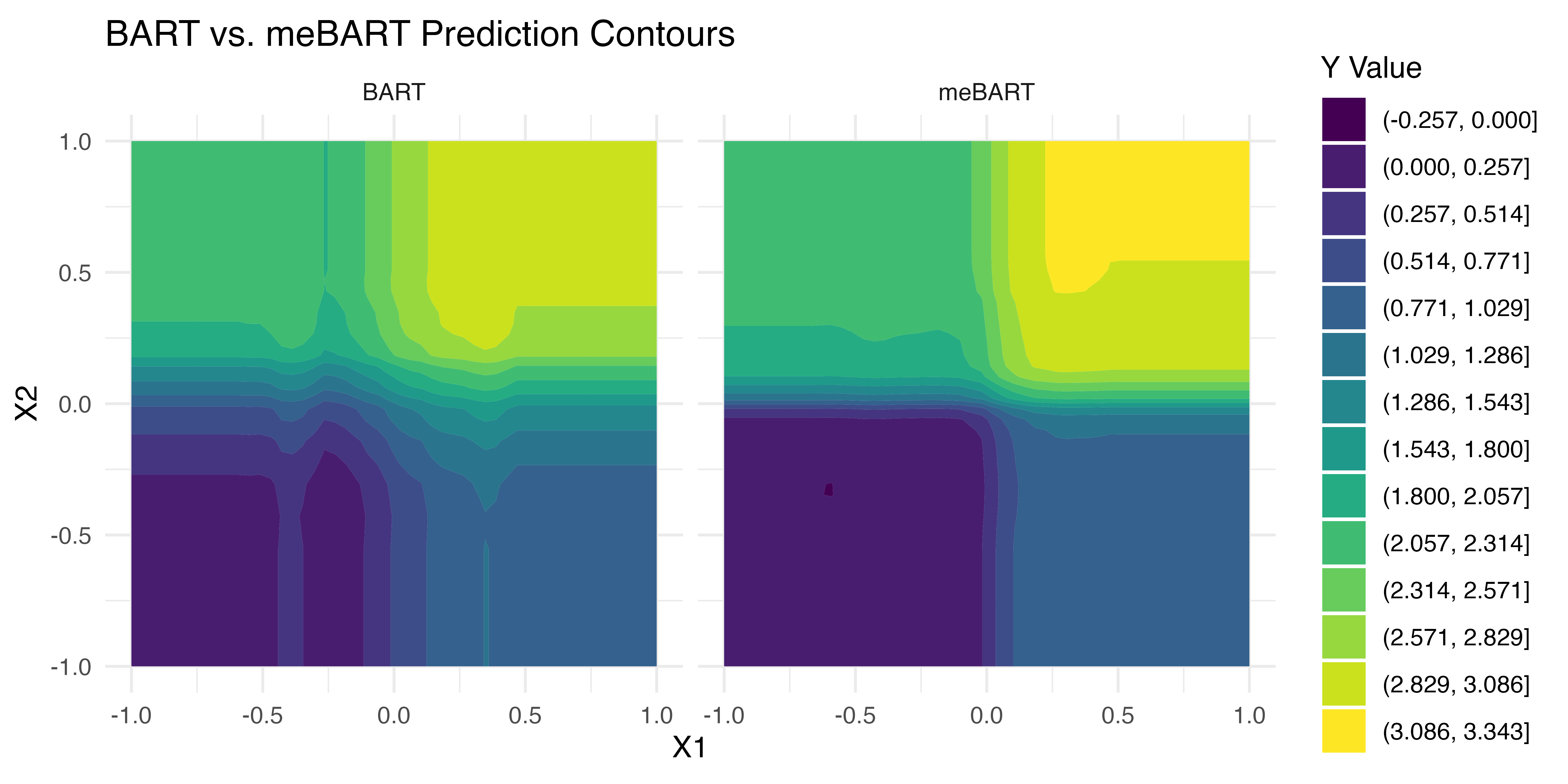}
     \caption{Contour plots of BART and meBART prediction values on synthetic data generated according to the 2-dimensional indicator function: $f(x_1, x_2) = \mathbbm{1}(x_1>0) + 2\times\mathbbm{1}(x_2>0)$.}
     \label{fig:2d-contours}
 \end{figure}

\end{document}